\documentclass[prl,floatfix,aps,twocolumn]{revtex4-1}

\usepackage[utf8]{inputenc}
\usepackage{graphicx}
\usepackage{amsmath}
\usepackage{amsfonts}
\usepackage{bm}
\usepackage{color}
\usepackage{subfigure}

\begin{document}
\title{Disordered multihyperuniformity derived from binary plasmas}

\author{Enrique Lomba$^{1,2}$, Jean-Jacques Weis$^3$, and Salvatore Torquato$^{2,4}$}
\affiliation{$^1$Instituto de Qu\'{i}mica F\'{i}sica Rocasolano,
CSIC, Calle Serrano 119, E-28006 Madrid,
Spain\\
$^2$Department of Chemistry, Princeton University, Princeton, New Jersey 08544, USA,\\
$^3$Université de Paris-Sud, Laboratoire de Physique Théorique, UMR8627, Bâtiment 210,
91405 Orsay Cedex, France\\
$^4$Princeton Institute for the Science and Technology of Materials, Princeton University, Princeton, New Jersey
08544, USA
}

\begin{abstract}

Disordered multihyperuniform many-particle systems are exotic amorphous 
states that allow exquisite color sensing capabilities due to their
anomalous suppression of density fluctuations for distinct 
subsets of particles, as recently evidenced in photoreceptor mosaics in avian retina.
Motivated by this biological finding, we present  the first statistical-mechanical
model that rigorously achieves disordered multihyperuniform
many-body systems  by tuning interactions
in binary mixtures of non-additive hard-disk plasmas.
 We demonstrate that multihyperuniformity competes with phase separation and
stabilizes a clustered phase. Our work provides a systematic means to generate
disordered multihyperuniform solids, enabling one to explore their potentially
novel photonic, phononic, electronic and transport properties.
\end{abstract}

\maketitle

A hyperuniform state of matter is characterized by an anomalous suppression of density
fluctuations at large length scales compared to the fluctuations in typical disordered point
configurations, such as atomic positions in ideal gases,
liquids, and glasses. A hyperuniform many-particle system in
$d$-dimensional Euclidean space $\mathbb{R}^d$ at number density $\rho$ is one
in which the structure factor $S({\bf Q})\equiv 1 + \rho {\tilde h}({\bf Q})$ tends to zero
as the wavenumber $Q \equiv |{\bf Q}|$ tends to zero \cite{To03a}, i.e.,
\begin{equation}
\lim_{Q\to 0} S({\bf Q}) = 0, 
\label{sq0}
\end{equation}
where ${\tilde h}({\bf Q})$ is the Fourier transform of the total correlation
function $h({\bf r}) = g_2({\bf r})-1$ and $g_2({\bf r})$ 
is the  pair correlation function.  Hyperuniform many-body systems include
crystals, quasicrystals, and certain exotic disordered systems \cite{To03a,Za09}. 
Disordered hyperuniform states lie between a crystal and liquid: they
behave like perfect crystals in the manner in which they suppress large-scale density
fluctuations and yet, like liquids and glasses, are statistically isotropic without Bragg
peaks \cite{To15}. Due to their  novel structural and physical properties, these  exotic states of amorphous matter have been the subject
of many recent investigations \cite{To08c,Za09,Fl09b,Za11a,Ji14,Ja15,To15,Ma15,To16b,Le16,Zh16b,Gh16,He17b,Gk17,Fr17,Ch18}.

Disordered {\it multihyperuniform} many-body systems are a remarkable
class of disordered hyperuniform states of matter that were first identified in the photoreceptor patterns in avian retina \cite{Ji14}. A disordered multihyperuniform point configuration
is not only disordered and hyperuniform  but contains multiple distinct subsets of the 
entire point configuration that  are themselves hyperuniform. This twist on standard hyperuniformity
is presumably partly responsible for the acute color sensing ability
of birds with their five different cone photoreceptor sub-populations, each of which
is separately hyperuniform. An open question since that time has been the identification of a theoretical
model of interacting particles
with  disordered multihyperuniform states.
 In this Letter, we provide the first statistical-mechanical
model of a many-body system that rigorously achieves disordered 
multihyperuniformity by tuning interactions
in binary mixtures of hard-disk plasmas.

Hyperuniform systems can be regarded to be at an ``inverted" critical 
point in which the direct correlation function $c({\bf r})$, defined
through the Ornstein-Zernike equation, is long-ranged and the total
correlation function $h({\bf r})$ is short-ranged, which is the diametric
opposite of the behaviors of these functions  at thermal critical points \cite{To03a}.
For a large class of disordered hyperuniform systems \cite{To16b}
\begin{equation}
  \lim_{Q\rightarrow 0} S(Q)\propto Q^\alpha
\end{equation}
with $\alpha > 0$. This can be thought of as an ``inverted" critical point 
and through the Ornstein-Zernike equation, this can be
shown to determine the low-$Q$ behavior of the Fourier transform of the direct correlation 
function $c({\bf r})$  \cite{To03a}:
\begin{equation}
   \lim_{Q\rightarrow 0} \tilde{c}(Q)\propto Q^{-\alpha}.
\label{C}
\end{equation}
Importantly,  at large pair separations, the direct correlation $c({\bf r})$  is equal to the negative of the
pair potential $u({\bf r})$ of a many-body system in equilibrium, and hence
the small-$Q$ behavior of the Fourier transform of the potential obeys the
following power-law form:
\begin{equation}
\lim_{Q\rightarrow 0} \tilde{u}(Q)\propto -Q^{-\alpha}.
\label{u}
\end{equation}
We see that condition (\ref{u}) dictates what type of interactions can 
lead to a hyperuniform system in equilibrium. Such a
situation is found for two- and three-dimensional Coulomb
plasmas\cite{PRB_1978_17_2827,PhysRep_1980_59_1,Ja81,Caillol1982,Levesque2000}.

However, to achieve multihyperuniformity, one must consider
particle mixtures, which is more
involved if the overall hyperuniformity condition (\ref{sq0})
is to be satisfied by the partial structure factors of each mixture component.
In this connection, it
was recently shown by the authors that superimposing a long-range Coulomb repulsion on a
non-additive hard disk (NAHD) mixture induces hyperuniformity only on a
global scale, and not for the structure of the mixture components \cite{Lo17}. 
In this Letter, we ascertain the conditions under which the
interactions can be tuned to induce
a multihyperuniform structure.

In general, in a binary mixture, the partial structure factors $S_{ij}(Q)$ can be
expressed in terms of the Fourier transform of the corresponding
total and direct correlation functions, $h_{ij}(r)$ and $c_{ij}(r)$, using the Ornstein-Zernike relation:
\begin{eqnarray}
  S_{ii}(Q)&=&x_i(1+\rho x_i\tilde{h}_{ii}(Q)) = x_i\frac{1-\rho_j\tilde{c}_{jj}}{|{\bf I -
      C}|} \nonumber \\
   S_{ij}(Q)&=& \rho x_ix_j\tilde{h}_{ij}(Q)  =  \frac{\rho x_ix_j\tilde{c}_{ij}}{|{\bf I -
      C}|} \label{ozmix}
\end{eqnarray}
with $i\neq j$, and 
\begin{equation}
  |{\bf I - C}| = 1 - \rho_1\tilde{c}_{11}
  -\rho_2\tilde{c}_{22}+\rho_1\rho_2(\tilde{c}_{11}\tilde{c}_{22}-\tilde{c}_{12}^2)\label{denom},
\end{equation}
where the $\rho_i$ are the partial number densities of each component,
$x_i$ the corresponding mole fractions, and $\rho=\rho_1+\rho_2$.
Intuitively, one can require each component interaction to display the
same singular behavior at $Q\rightarrow 0$ as found in disordered
single component hyperuniform systems, namely $\lim_{Q\rightarrow 0}
\tilde{c}_{ij}(Q) \sim -\beta \tilde{u}_{ij}(Q)\propto -\eta_{ij}
Q^{-\alpha}$. Here $\eta_{ij}$ is a constant dependent on the type of
interaction and $\beta=1/(k_BT)$ as usual. For standard Coulomb interactions, one has
$\eta_{ij}\propto z_iz_j$, by which
$\eta_{ij}=\sqrt{\eta_{ii}\eta_{jj}}$. In contrast to the single-component 
case, one observes that 
 the denominator (\ref{denom}) contains a quadratic term that, in
principle, can guarantee that $\lim_{Q\rightarrow 0}S_{ij}=0$, $\forall
(i,j)$, if 
\begin{equation}
  \lim_{Q\rightarrow
    0}\beta^2\left(\tilde{u}_{11}(Q)\tilde{u}_{22}(Q)-\tilde{u}_{12}(Q)^2\right)
  \neq 0.
  \label{berth}
  \end{equation}
Here we have made use of the large-$r$ asymptotic behavior of the direct
correlation function $c_{ij}(r)$.  If
instead, the equality in  expression (\ref{berth}) - basically one of
Lorentz-Berthelot's mixing rules - is 
fulfilled, then Eqs.~(\ref{ozmix}) lead to non-vanishing partial
structure factors for $Q\rightarrow 0$, even if the total structure
factor complies with (\ref{sq0}) \cite{Lo17}. In what follows, we
will show for a simple mixture model in two dimensions how
multihyperuniformity can be achieved by a simple tuning of the cross
interaction long-range behavior so that it satisfies the inequality (\ref{berth}).

Our model consists of a symmetric mixture of non-additive hard disks
with a two-dimensional Coulomb repulsion added, such that the pair potential is given by
\begin{equation}
  \beta u_{ij}(r) = \left\{
  \begin{array}{cc}
    \infty & r < (1+\Delta(1-\delta_{ij}))\sigma \\
    - \gamma_{ij}z_iz_j\Gamma \log r/\sigma
    \label{uijlog} &  r \ge (1+\Delta(1-\delta_{ij}))\sigma
  \end{array}
  \right.
\end{equation}
 where $\gamma_{ij}=(\lambda + (1-\lambda)\delta_{ij})$, $\Gamma =\beta e^2$, being $e$ the
 electron charge (esu units), $z_i$, the particle charge in $e$ units,
 $\delta_{ij}$ is a Kronecker $\delta$, $\Delta$ is the
 non-additivity 
 parameter, and  $\sigma$ the hard disk diameter between like
 species. For a plasma system we consider $z_i>0$ for all species. The
 key element in Eq.~(\ref{uijlog}) is the $\lambda$ 
 parameter, which can be tuned in the range $0\le \lambda\le 1$. When
 $\lambda=1$ we have the regular situation in which $u_{12} \propto
 z_1z_2$. As $\lambda < 1$, the unlike interaction become less
 repulsive, to go back to the plain hard-core repulsion when
 $\lambda=0$. In all cases, when $\lambda\neq 1$, inequality (\ref{berth}) is 
 fulfilled.  Finally, our system is taken to be a symmetric mixture of
 total number density, 
 $\rho$, with $\rho_1=\rho_2=\rho/2$ and $z_1=z_2$. 

\begin{figure}
  \includegraphics[width=7cm,clip]{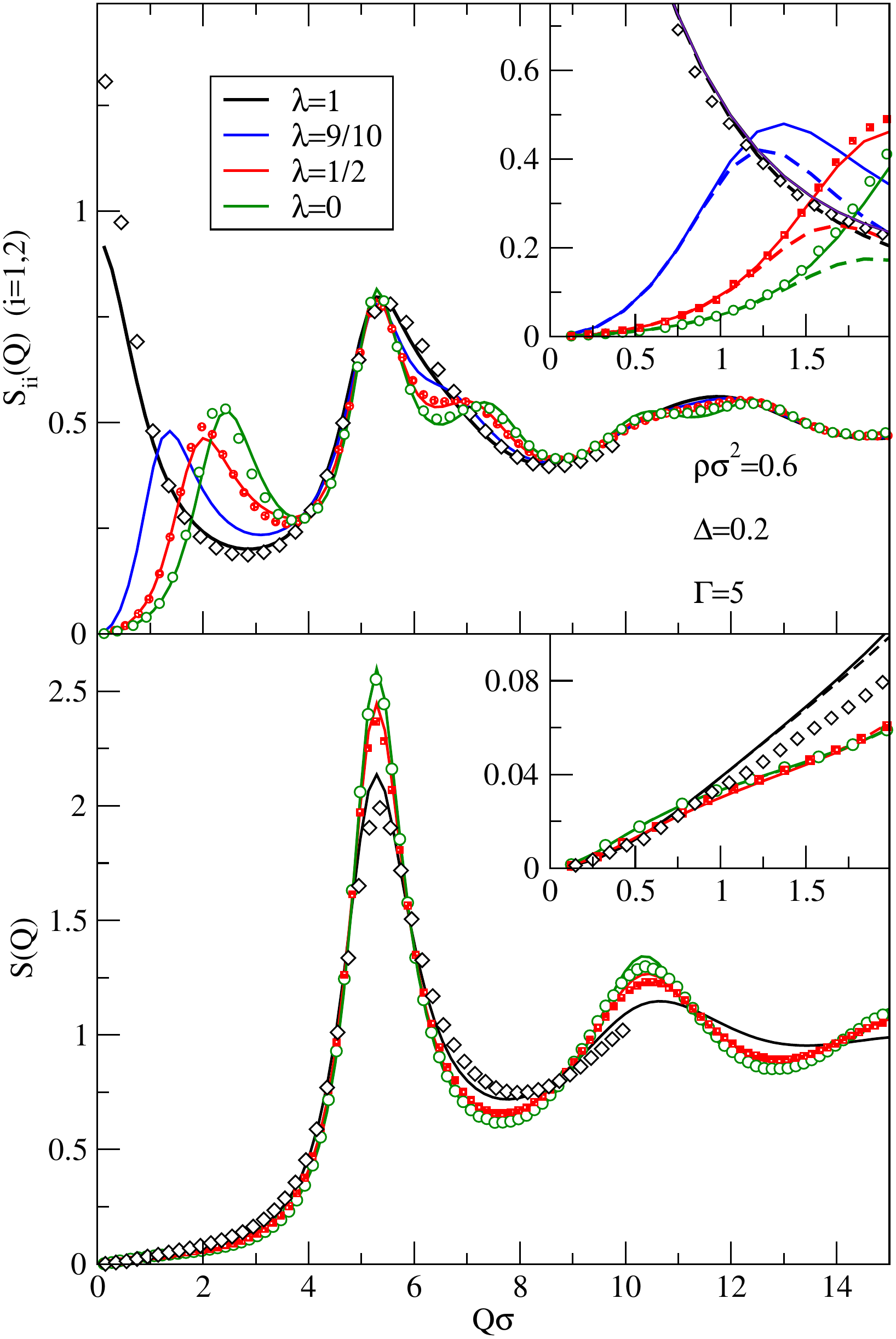}

\caption{Total and partial structure factor of the NAHD plasma for
  various values of the $\lambda$ parameter tuning the Coulomb
  interaction between unlike particles. Solid curves correspond to RHNC-PY calculations, symbols denote Monte Carlo data. The multihyperuniformity
  induced by Coulomb interactions when $\lambda \ne 1$ counteracts
  phase separation by  
  stabilization of a cluster-like phase with intermediate range
  order. This is reflected in the  prepeak occurring at $Q\sigma \lesssim
  2.3$\label{sqvl}. The insets illustrate the low-$Q$ behavior of
  $S_{ii}(Q)$ as determined from Eq.(S3) in the SI, and of
$S(Q)$ from Eq.(\ref{snnl}) --dashed curves--. Note that given the
  symmetry in the interactions and composition, $S_{11}=S_{22}$.}
\end{figure}

In order to analyze the low-$Q$ behavior of the partial structure
factors, we 
perform a small-$Q$ expansion of ${\tilde c}_{ij}$ and  separate out the Coulomb term:
\begin{eqnarray}
  \tilde{c}_{ij}(Q) &\approx& \tilde{c}^R_{ij}(0)+c_{ij}^{(2)}Q^2- \gamma_{ij} 2\pi\Gamma
  z^2/Q^2\nonumber\\
  c_{ij}^{(2)}&=&\frac{1}{2}\left.\frac{\partial \tilde{c}_{ij}^R
    (Q)}{\partial Q}\right|_{Q=0},
  \label{limcq}
\end{eqnarray}
where,  $c_{ij}^R(0)$ is  the $Q\rightarrow 0$ limit of the regular part of the direct correlation
function \cite{PhysRep_1980_59_1}. Inserting (\ref{limcq}) into
Eq.~(\ref{ozmix}),  the low-$Q$ limits of the partial structure
factors are obtained. With these, the total
structure factor, which in our  symmetric case reduces to  
\begin{equation}
  S(Q) = 2S_{11}(Q)+2S_{12}(Q),
\end{equation}
can be expressed as
\begin{equation}
  S(Q) \approx \frac{Q^2}{aQ^2+\pi\rho\Gamma z^2 (1+\lambda)}
  \label{snnl}
 \end{equation}
  when $Q\rightarrow 0$.
The concentration-concentration structure factor\cite{Bhatia1970} is
defined in terms 
of the Fourier transform of the local concentration deviation from its
average value, $\tilde{C}(Q)$, as
\begin{eqnarray}
S_{cc}(Q) &=& N\langle \tilde{C}({\bf Q})\tilde{C}(-{\bf
  Q})\rangle\nonumber\\
&=& x_2^2S_{11}(Q)+x_1^2S_{22}(Q)-2x_1x_2S_{12}(Q)\nonumber
\end{eqnarray}
where, $N$ in the number of particles, and $\langle\ldots\rangle$ denotes an
ensemble average. In our fully symmetric equimolar case we have
\begin{equation}
  S_{cc}(Q)  =  \frac{1}{2}(S_{11}(Q)-S_{12}(Q)),
\end{equation}
which leads to ($Q\rightarrow 0$)
\begin{equation}
  S_{cc}(Q) \approx \frac{Q^2}{bQ^2+\pi\rho\Gamma z^2 (1-\lambda)}.
  \label{sccl}
\end{equation}
In (\ref{snnl}) and (\ref{sccl})
\begin{eqnarray}
  a & = & 1 -\frac{\rho}{2}(\tilde{c}_{11}^R(0)+\tilde{c}_{12}^R(0)+(c_{11}^{(2)}+c_{12}^{(2)})Q^2)\nonumber
  \\
  b & = & 1 -\frac{\rho}{2}(\tilde{c}_{11}^R(0)-\tilde{c}_{12}^R(0)+(c_{11}^{(2)}-c_{12}^{(2)})Q^2).\nonumber
\end{eqnarray}
These simple expressions (\ref{snnl}) and (\ref{sccl}) are valid for
all values of $\lambda$, and can be seen to accurately
reproduce the low-$Q$ behavior of the structure factors in the
insets of  Figure \ref{sqvl}. When $\lambda=1$ (usual Coulomb
charge-charge interaction) one sees that $S(Q)\propto Q^2$ and
$S_{cc}(Q)\propto 1/b$ as $Q\rightarrow 0$, i.e., the system will only
be globally hyperuniform. For all other cases, the total and
concentration-concentration structure factors vanish simultaneously as
$Q\rightarrow 0$, meaning that the system is necessarily
multihyperuniform. This is a general result: provided that the
long-range part of the interaction satisfies Eq.~(\ref{u}) and inequality
(\ref{berth}) holds, the system will display hyperuniformity. Neither
the interaction symmetry implicit in Eq.(\ref{uijlog}) nor the
equimolar composition are essential to our results. On the other hand,
if interactions do not comply with (\ref{berth}), global
hyperuniformity will only occur if the symmetry $\lim_{Q\rightarrow
  0}(\tilde{u}_{11}(Q)-\tilde{u}_{22}(Q))=0$ is preserved\cite{Lo17}.

 \begin{figure*}
{\includegraphics[width=7cm,clip]{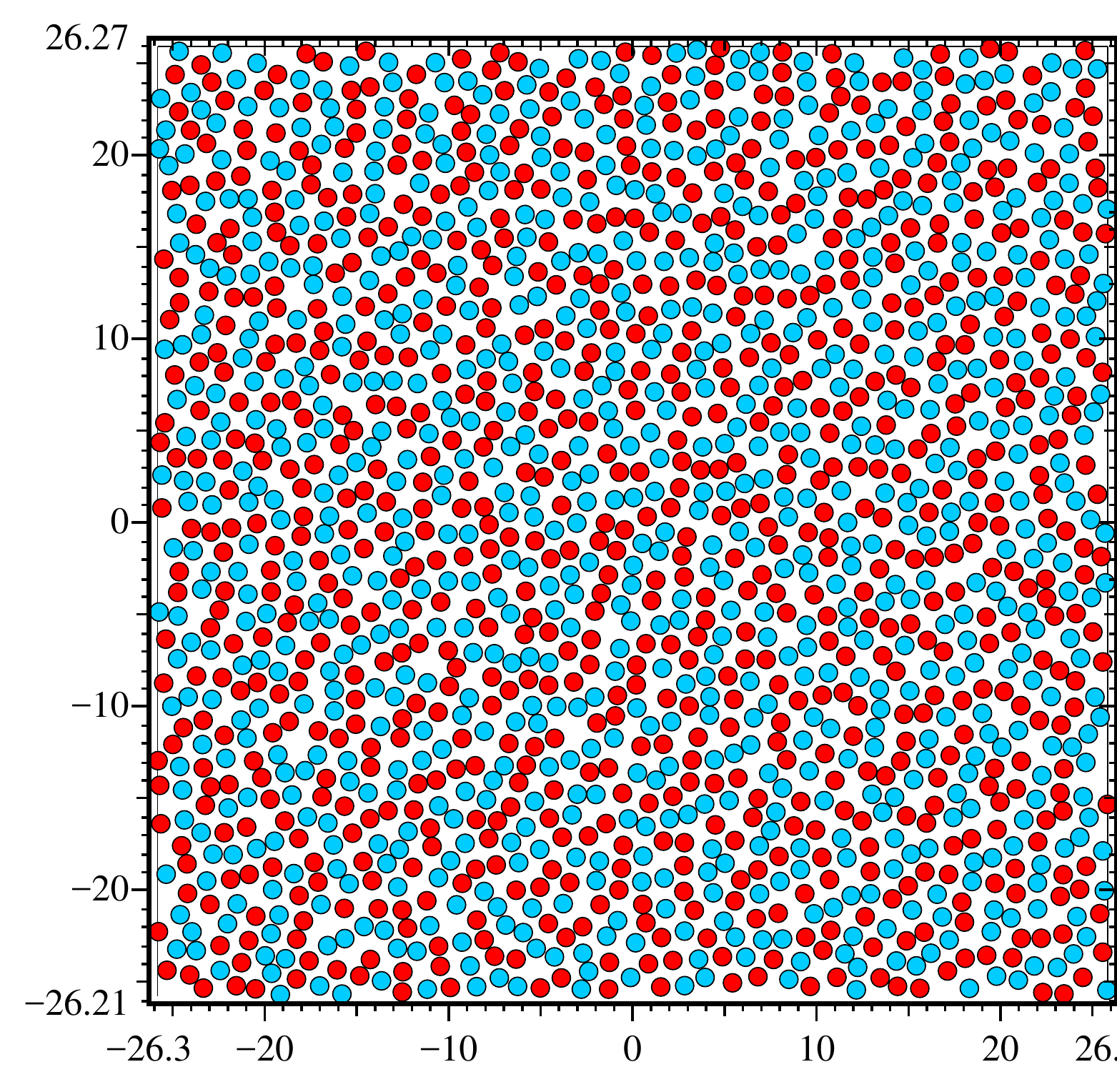}\label{snapl0}}{\includegraphics[width=7cm,clip]{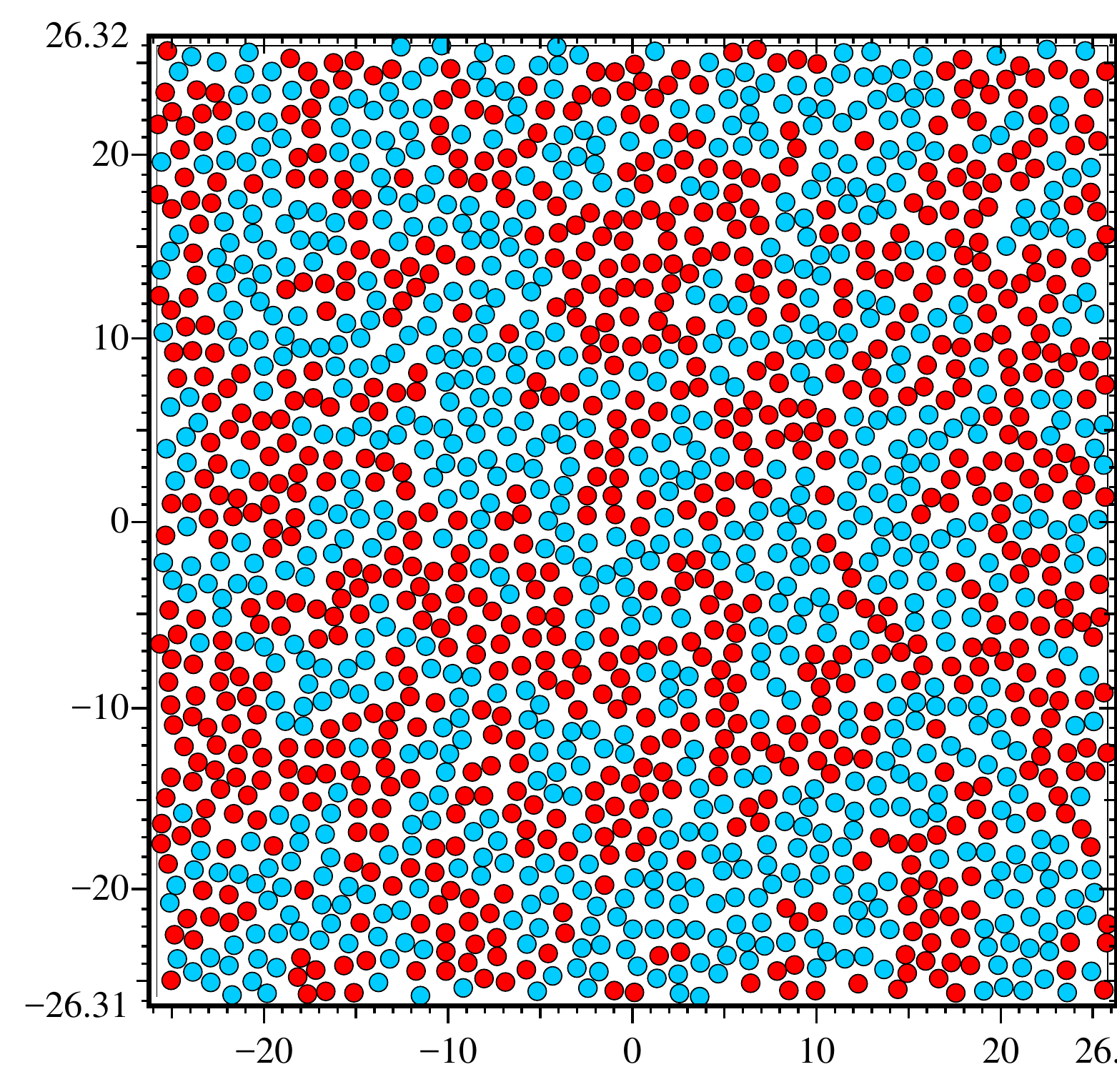}\label{snapl10}}
   \caption{Snapshots of Monte Carlo configurations of the NAHD two-
component plasma with the unlike Coulomb interaction turned off
($\lambda=0$, left panel), and identical Coulomb interaction
     between like and unlike particles ($\lambda=1$, right panel). One sees the build up of 
     mostly linear finite clusters of particles in the
     multihyperuniform case, i.e. when the unlike Coulomb interaction
     is turned off ($\lambda=0$).} 
\label{snap}
   \end{figure*}

We have studied our system resorting to the solution of the
Reference Hypernetted Chain Equation (RHNC), with a bridge function
of a plain NAHD reference fluid  obtained in the
Percus-Yevick (PY) approximation. Specific numerical 
procedures are detailed in
Ref. \cite{JCP_2007_127_074501}. Additionally, we have made extensive
use of Monte Carlo simulations, in which long-range interactions have
been treated using Ewald summations with conducting boundary
conditions \cite{JCP_2007_127_074501}; see Ref. \cite{Lo17}
for additional details.

A particular interesting situation occurs when one deals with positive
non-additivity. This is a well known situation in which volume effects
in the NAHD system will lead to
demixing \cite{Saija2002,Almarza2015}. Here we will focus on a system
with $\rho\sigma^2=0.6$, $\Delta=0.2$  and $\Gamma=5$ (relatively close to
the demixing critical point, located at $\rho_c\sigma^2=0.69$
\cite{Lo17}). Interestingly, according to
Eq.~(\ref{sccl}), once $\lambda\ne 1$, long-range concentration-concentration
fluctuations will be suppressed (the system becomes multihyperuniform)
and hence phase separation is no longer possible. This effect is
illustrated in Figure \ref{sqvl}, where the evolution of the partial,
$S_{ii}(Q)$,  
($i=1,2$) and the total structure factor, $S(Q)$, is displayed for various
$\lambda$ values. For $\lambda=1$, we see that $S_{11}$ grows rapidly
as $Q\rightarrow 0$ (in parallel with $S_{cc}(Q)$, see Supplementary Information), a clear indication
of the vicinity of the demixing transition. Then, for $\lambda\neq 1$,
one observes the presence of a prepeak whose 
position in $Q$-space increases as $\lambda$ decreases. This prepeak is the
characteristic signature of intermediate range
order \cite{Godfrin2013}, and it usually reflects spatial correlations
between stable clusters. Indeed, the effect of lowering
$\lambda$ can be readily interpreted in terms of competition between
effective interactions. The volume effects induced by the positive
non-additivity are equivalent to effective short-range attractions
between like particles in the mixture. These effective short-range
attractions are responsible for the demixing that takes place when
density is increased \cite{Almarza2015}. Now, lowering $\lambda$ leaves
a long-range repulsion between like particles that is no longer
compensated by a similar repulsion between unlike particles. As a
result, we have now a system that exhibits a short-range attraction
and long-range repulsion (SALR) between like particles,  a
well-known class of potentials characterized by inducing the presence
of stable clusters (as well  lamellar and bicontinuous)
phases \cite{Imperio2006,Imperio2007}. For simple fluids,
the presence of long-range repulsions is known to inhibit the
liquid-vapor transition in favor of the formation of a cluster
phase \cite{Godfrin2013,Godfrin2014}. Our situation is completely
analogous, with the uncompensated long-range Coulombic repulsion
between like particles 
inhibiting the demixing transition, stabilizing the transient particle clusters
that occur for $\lambda=1$ close to the consolute point, and thus leading
to the presence of a prepeak in $S_{ii}(Q)$. Moreover, these
uncompensated Coulomb repulsions turn
the system multihyperuniform, as seen in
the low-$Q$ behavior of $S_{ii}(Q)$ and $S(Q)$ depicted in
Figure \ref{sqvl}.  The clustering effect is readily seen in the
snapshots shown in Figure \ref{snap} for $\lambda=1$ and
$\lambda=0$. Observe that the multihyperuniform configuration
($\lambda=0$) is mostly dominated by 
the presence of linear particle clusters (chains) of finite size, in contrast with the
more extended clusters found for $\lambda=1$, which is
only globally hyperuniform. The multihyperuniform systems exhibit
microsegregation with stable clusters, whereas for $\lambda=1$
clusters correspond to transient states on the way to the complete phase
separation that occurs once the total density is increased by a
small amount. This 
microsegregation (or microheterogeneity) is characteristic of the
presence of SALR-type interactions. The average chain-chain separation
for $\lambda=0$ 
can be estimated from the position of the prepeak in $S_{11}(Q)$,
$Q_m\approx 2.3\sigma^{-1}$, as $2\pi/Q_m\approx 2.73\sigma$. Visual
inspection of the configuration in Figure \ref{snapl0} is in accordance with
this estimate. Basic energetic considerations can explain the
compositional order of Figure \ref{snapl0}. When $u_{AB}^{LR} <
u_{AA}^{LR}=u_{BB}^{LR}$, the simplest way to 
minimize repulsion between like particles is a chain-like
configuration. At the same time, chains of A(B) particles would tend
to be surrounded basically by chains of unlike, B(A) particles, with
a smaller repulsion (just the hard-core when $\lambda=0$). Results for the cluster
size distributions and number density fluctuations are given in the Supplementary Information.

In summary, we have presented  the first statistical-mechanical
model that rigorously achieves disordered multihyperuniform
many-body systems  by tuning interactions
in binary mixtures of non-additive hard-disk plasmas.
Interestingly, multihyperuniformity competes with
phase separation and stabilizes a clustered phase.

The exploration of the potentially novel properties
of multihyperuniform materials is a wide open research area.
Standard disordered hyperuniform solids  have already been shown to be  of importance because they have novel 
photonic, phononic, electronic and transport properties \cite{Fl09b,Xie13,Le16,Zh16b,Fr17,Gk17,Ch18},  despite the lack of translational order. 
Using designer material techniques, examples have been 
fabricated in the laboratory on the microwave scale and their photonic properties have been studied 
\cite{Man13b}.  Our work enables one to 
generate disordered multihyperuniform dispersions, and then
computationally determine their physical properties. Such promising designer materials
could then be fabricated using 3D printing technologies.

E. L.  acknowledges the support from the Direcci\'on
General de Investigaci\'on Cient\'{\i}fica  y T\'ecnica under Grant
No. FIS2013-47350-C5-4-R, and  from the Program Salvador de
Madariaga, PRX16/00069 which supports his sabbatical stay at Princeton
University.
S. T. was supported by
the National Science Foundation under Award No. DMR-
1714722.


%
\end{document}